%% file: coriandolib.tex
\documentclass[aps,prl,twocolumn]{revtex4}  \usepackage{graphicx}
\usepackage{amsmath}
\input psfig2.tex
\begin{document}

\title{Geometric phase in external electromagnetic fields}

\author{Simone Selenu}
\affiliation{Atomistic Simulation Center, School of Mathematics and
 Physics\\ Queen's University Belfast, Belfast BT7 1NN, Northern
 Ireland, UK}

\begin{abstract}
Variations of $\it{polarization}$ is a dielectric property of 
quantum systems, quantified by Resta et al. \cite{Resta2}, and discovered to
be a Berry phase\cite{Berry} of the electronic subsystem. 
In order to continue the previous research we wrote a scalar phase $\Phi$ as
the $\it{circulation}$ of the polarization field,a  vector field called
$\it{dipole~moment~per~unit~volume}$ $\bar{P}$, or in alternative and in a
more general form as the flux of a $\it{new}$
  vector field called $macroscopic~rotonic~field$ indicated with
  ${\bf{L^{+}}}$. The latter shows to be a more general berry phase concerning the 
 interaction of the electronic subsystem with an external electromagnetic field.
 
\end{abstract}
 
\date{\today} \maketitle

\section{Introduction}
\label{introduction}

In this paper is developed a model in order to introduce a general definition
of a Berry phase\cite{ZakBerry,Zak9} concerning the interaction of the
electronic subsystem with  a uniform static electromagnetic 
field. The latter phase is
what I call a scalar quantum phase, a Berry phase. 
This heuristic study is based on mechanical and electrical properties of the 
electronic field in its steady states\cite{coriandoli}. 
Firstly, we introduce several equations useful for the derivation of 
the scalar phase when only a uniform electrostatic field is interacting with the
electronic field, we then generalize the result considering also an external
 magnetic
field interacting with the electronic subsystem. A derivation of two
$\it{new}$ vectorial fields will be given and their relation to geometric
phases will be exploited.

\section{Group velocity in finite electrostatic fields}
In this section we derive heuristically or better theoretically the expression of the 
$\it{group~velocity}$ of the electronic field as a first step of our
understanding of the dielectric responce of the latter when it is interacting
with an external electrostatic field.
We make use of a particular $\it{gauge~representation}$, the $k-q$ 
representation\cite{Zak}. We always considered the eigenstates of the system as follows:

\begin{equation}
\Psi_{n,\bf{k}} = e^{i\bf{k}\cdot\bf{q}} u_{n,\bf{k}}
\end{equation}

and the system being in its steady states\cite{coriandoli}. The Hamiltonian is
written as follows:

\begin{equation}
\label{hamilt}
\begin{split}
H_{{\bf{k}}} \equiv &[\frac{1}{2m}(-i\hbar \nabla_{{\bf{q}}}+{\hbar
  \bf{k}})^{2}- {\bf{E^{0}}}\cdot i\nabla_{{\bf{k}}} +
  V({\bf{q}})]
\end{split}
\end{equation} 

where ${\bf{E^{0}}}$ is a uniform static external electric field.

It is already known that the expectation values of the $\it{group~velocity}$
operator\cite{coriandoli}:

\begin{equation}
\label{grvel}
{\bf{\hat{v}}}=\frac{1}{m}(-i\hbar\nabla_{{\bf{q}}}+ \hbar{\bf{k}}) 
\end{equation}

 in a vanishing external electromagnetic field 
are written as follows:

\begin{equation}
\label{jmn}
{\bf{v}_{mn}}=<u_{m,{\bf{k}}}|{\bf{\hat{v}}}|u_{n,{\bf{k}}}> 
\end{equation}

The same result holds when a uniform static electric field is present 
in the system by direct application of theorem in\cite{coriandoli} to the Hamiltonian 
(\ref{hamilt}). 
In absence of external fields it has been shown\cite{ZakBerry} that:

\begin{equation}
\label{jmn1}
{\bf{v}_{mn}}=-i {\bf{}
{\bf{} \omega_{mn}}}{\bf{d}_{mn}} + \frac{1}{\hbar}
\nabla_{{\bf{k}}}\epsilon_{n,{\bf{k}}} {\bf{} \delta_{mn}}
\end{equation} 
 previous result $\it{holds}$ also for any applied external
 uniform electrostatic field present in the system. 
Transition frequencies are written as: 
${\bf{} \omega_{mn}}=\frac{\epsilon_{m,{\bf{k}}}-\epsilon_{n,{\bf{k}}}}{\hbar}$
and  ${\bf{v}_{mn}}$ are matrix elements.
Matrix elements ${\bf{d}_{mn}}$ are instead the matrix elements of the $\it{position}$
operator, and are defined as:

\begin{equation}
\label{Pmn0a}
{\bf{d}_{mn}}=<u_{m,{\bf{k}}}| i\nabla_{{\bf{k}}}|u_{n,{\bf{k}}}>
\end{equation} 

and because of the normalization condition on the amplitudes $u_{n,\bf{k}}$ follows that:

\begin{equation}
\label{Pmn1a}
{\bf{d}^{*}_{mn}}= {\bf{d}_{nm}}
\end{equation} 

implying that diagonal fields ${\bf{d}_{nn}}$ are real, and the
matrix $ {\bf{d}_{nm}}$ is Hermitian. 
It is possible to show to be valid the following relations:

\begin{equation}
\label{jmn2}
\begin{split}
{\bf{d}_{mn}} &= i\frac{{\bf{v}_{mn}}}{{\bf{} \omega_{mn}}} ~~  m \neq
n \\ {\bf{v}_{nn}} &= \frac{1}{\hbar}
\nabla_{{\bf{k}}}\epsilon_{n,{\bf{k}}} ~~  m =  n \\
\end{split}
\end{equation} 

as previously shown by\cite{ZakBerry} when vanishing external fields are concerned. It is remarkable that the second
definition, of eq.(\ref{jmn2}), can be directly derived by the theorem stated 
in\cite{coriandoli}.
It represents the $\it{group~velocity}$\cite{coriandoli} of the electronic
subsystem. Armed with the previous result we derive\cite{ZakBerry} a Berry phase of the
electronic system in the following section.

\section{Scalar phase in eternal electrostatic fields}
Here we derive a Berry phase of the electronic field bearing in mind that 
the Hamiltonian of the system is reported in eq.(\ref{hamilt}).
As shown by Zak and Berry\cite{Berry,Zak9}, we can write:

\begin{equation}
\label{flusso}
\begin{split}
\int_{S} d{\bf{S}} \cdot {\bf{Im}} \sum_{m \neq n}
[\frac{{\bf{v}_{mn}} \times {\bf{v}_{nm}}}{\omega^{2}_{mn}}]
=\int_{S} d{\bf{S}} \cdot \nabla_{\bf{k}} \times {\bf{d}_{nn}}\\
\end{split}
\end{equation}

where $S$ is a closed surface of the system, Brillouine zone, and $C$ is a
curve contained in that surface. 
Also,

\begin{equation}
\label{flusso1}
\begin{split}
\int_{S} d{\bf{S}} \cdot \nabla_{\bf{k}} \times {\bf{d}_{nn}}
=\oint_{C} {\bf{d}_{nn}} \cdot {d\bf{l}}\\
\end{split}
\end{equation}

and finally,

\begin{equation}
\label{flusso2}
\begin{split}
\oint_{C} {\bf{d}_{nn}} \cdot {d\bf{l}}=\phi_{n} \\
\end{split}
\end{equation}

On the other hand, we can write\cite{Berry,Zak9}:

\begin{equation}
\label{uno}
\int_{S}d\bf{S}\cdot Im\sum_{\bf{m \neq n}} \bf{d_{mn}} \times  \bf{d_{nm}}= 
\int_{S}d\bf{S}\cdot \nabla_{k} \times  \bf{d_{nn}}
\end{equation}

It is then possible to express $\phi_{n}$ in another way:

\begin{equation}
\label{due}
\int_{S}d\bf{S}\cdot Im\sum_{\bf{m \neq n}} \bf{d_{mn}} \times  \bf{d_{nm}}= \phi_{n}
\end{equation}

where $\bf{Im}$ stands for imaginary.

After a summation over $n$, it is possible to evaluate:

\begin{equation}
\label{sum-flusso}
\sum_{n} \phi_{n} = \Phi
\end{equation}

where $\Phi$ is a $\it{conserved~ geometric~ phase}$ (a Berry phase). We shall call
$\Phi$ a first scalar phase\cite{Berry}. 
The averaged trace of fields ${\bf{d}_{mn}}$ can be
calculated by integration over the Brillouin zone in reciprocal space as
and gives the macroscopic polarization\cite{Resta2,KS-V}:

\begin{equation}
\label{DefP}
\begin{split}
\bar{{\bf{P}}} =\frac{e}{(2 \pi)^{3}} \int_{BZ} dk {\bf{}Tr}({\bf{d}_{mn}})
\end{split}
\end{equation}

It is possible to define a $\it{dipolar~field}$ as follows:

\begin{equation}
\label{DfP}
\begin{split}
{\bf{P}} =e{\bf{}Tr}({\bf{d}_{mn}})
\end{split}
\end{equation}

Making use of equations (\ref{flusso2})(\ref{sum-flusso})(\ref{DfP})allows us to define the scalar phase as follows:

\begin{equation}
\label{Berryu}
\begin{split}
\oint_{C} {\bf{P}} \cdot {d\bf{l}} =e\Phi \\
\end{split}
\end{equation}
 
where $C$ is a closed curve contained in the surface of the Brillouin
zone. Berry phase instead is given by eq.(\ref{Berryu}) and quantifies the
amount of polarization charge  $e\Phi$ $\it{stored}$ on the surface of the
system (a unit cell if the system is a crystal\cite{Ashcroft}).
We may calculate the conserved phase $\Phi$, not only by
the knowledge of matrix elements of the velocity operators, 
as in eq.(\ref{flusso}),(\ref{flusso1}),(\ref{flusso2}), but also by the knowledge of matrix elements 
${\bf{d}_{mn}}$ eq.(\ref{due}). Result of eq.(\ref{Berryu}) is also in
agreement with the modern theory of polarisation where the phase $\Phi$ is
proportional to the 
Chern number of a Chern-insulator\cite{KS-V1,ResV1}, we still consider it either a 
Berry or a Zak phase calling it a scalar geometric phase.
In the following, we extend the problem to find a scalar phase proportional to
the Chern number to the case of an external
uniform magnetostatic field interacting with the electronic field.

\section{Scalar phase in electromagnetic fields}
\label{Vortex}

In this section we generalize the problem to the case of  also an external
uniform magnetostatic field $B^{0}$ interacting with the electronic system. 

We can write the Hamiltonian of the system\cite{Zak} as follows:

\begin{equation}
\label{vel-kq6}
\begin{split}
H_{{\bf{k}}} \equiv &[\frac{1}{2m}(-i\hbar \nabla_{{\bf{q}}}+{\hbar
  \bf{k}}-\frac{e}{2c} B^{0} \times i\nabla_{{\bf{k}}})^{2} -E_{0}\cdot
 i\nabla_{{\bf{k}}} + V({\bf{q}}) ]
\end{split}
\end{equation} 

where $e$ is the elementary electronic charge.

In analogy to eq. (\ref{jmn}) we can define the following matrix:

\begin{equation}
\label{jmn-B01}
\begin{split}
{\bf{v^{(B^{0})}}}_{mn}&=-i {\bf{} \omega_{mn}} {\bf{d}_{mn}}  + {\bf{V}_{mn}}  \\&+ \frac{e}{2mc}
B^{0} \times {\bf{d}_{mn}}\\
\end{split}
\end{equation} 

where,

\begin{equation}
\label{jmn-B012}
{\bf{V}_{mn}}=\frac{1}{\hbar}\nabla_{{\bf{k}}}\epsilon_{n} {\bf{} \delta_{mn}}  
\end{equation}  

We shall see that the scalar phase introduced in eq.(\ref{Berryu}) can be
directly obtained as the $\it{flux}$ over a closed surface of a new vectorial field.
Let us define:

\begin{equation}
\label{PxJ1}
\begin{split}
{\bf{l}^{+}_{n}}&=\frac{1}{2}\sum_{m \neq n }\frac{[{\bf{d}_{nm}} \times
   {\bf{v}^{(B^{0})}_{mn}}] + [{\bf{d}_{mn}} \times
   {\bf{v}^{(B^{0})}_{nm}}]}{\omega_{mn}} \\&= {\bf{rot}~{\bf{d}}_{nn}}\\
\end{split}
\end{equation}

where ${\bf{rot}~{\bf{d}}_{nn}}=\nabla_{k} \times  \bf{d_{nn}}$, so that integrating over the  surface of the Brillouin zone we obtain:  

\begin{equation}
\label{conserved}
\sum_{n}\int_{BZ} d\bf{S}\cdot rot\bf{d_{nn}} = \sum_{n}\int_{BZ} d\bf{S}\cdot
\bf{l}^{+}_{n}
\end{equation}

It is worth noticing here that the vectorial field $\bf{l}^{+}_{n}$ 
relates the matrix elements of the velocity operator to the matrix elements of
the position operator of the electronic field. Note
also that the sum in eq.(\ref{PxJ1}) is weighted by the factors $\frac{1}{\omega_{mn}}$.
We can then define a new macroscopic vectorial field as follows:

\begin{equation}
\label{L}
{\bf{L}^{+}}=\sum_{n}{\bf{l}^{+}_{n}}
\end{equation} 

and call it $\it{rotonic~field}$. Also, in analogy to
eq.(\ref{uno}),(\ref{due}),(\ref{sum-flusso}) we can define a Berry
phase as follows:

\begin{equation}
\label{conserved1}
\Phi^{B^{0}}= \int_{BZ} d\bf{S}\cdot \bf{L}^{+} 
\end{equation}

expressible as the flux of the rotonic field. 

It is clear that eq.(\ref{conserved1}) is the more general expression of the 
Berry phase $\Phi$. In fact, in absence of magnetic field the latter
expression  (\ref{conserved1}) reduces to eq.(\ref{flusso}) once we substitute
matrics elements $\bf{d_{mn}}$ of eq.(\ref{jmn2}) in eq.(\ref{PxJ1}).

\section{Summary}
A Berry phase $\Phi$, directly dependent on the dielectric
properties of the electronic subsystem because of its relation with the
polarization  field
can be expressed, in a purely mechanical way, as the ${\bf{flux}}$ of the
$\it{rotonic~field}$, ${\bf{L^{+}}}$, over a closed surface in the Brillouin
zone allowing for the definition of the Chern number of the electronic field
interacting with an external finite static electromagnetic field.

My acknowledgment to Prof. Fiorentini and Dr. Colizzi for useful
discussions. I would like to acknowledge Atomistic Simulation Centre (ASC) for
the pleasent research.

This work has been supported by the ESF grant.

\bibliography{refs}

\end{document}

%% file: psfig2.tex
\def\PsfigVersion{1.9}
\ifx\undefined\psfig\else \fi

\let\LaTeXAtSign=\@
\let\@=\relax
\edef\psfigRestoreAt{\catcode`\@=\number\catcode`@\relax}
\catcode`\@=11\relax
\newwrite\@unused
\def\ps@typeout#1{{\let\protect\string\immediate\write\@unused{#1}}}
\ps@typeout{psfig/tex \PsfigVersion}

\def\figurepath{./}

\def\@nnil{\@nil}
\def\@empty{}
\def\@psdonoop#1\@@#2#3{}
\def\@psdo#1:=#2\do#3{\edef\@psdotmp{#2}\ifx\@psdotmp\@empty \else
    \expandafter\@psdoloop#2,\@nil,\@nil\@@#1{#3}\fi}
\def\@psdoloop#1,#2,#3\@@#4#5{\def#4{#1}\ifx #4\@nnil \else
       #5\def#4{#2}\ifx #4\@nnil \else#5\@ipsdoloop #3\@@#4{#5}\fi\fi}
\def\@ipsdoloop#1,#2\@@#3#4{\def#3{#1}\ifx #3\@nnil 
       \let\@nextwhile=\@psdonoop \else
      #4\relax\let\@nextwhile=\@ipsdoloop\fi\@nextwhile#2\@@#3{#4}}
\def\@tpsdo#1:=#2\do#3{\xdef\@psdotmp{#2}\ifx\@psdotmp\@empty \else
    \@tpsdoloop#2\@nil\@nil\@@#1{#3}\fi}
\def\@tpsdoloop#1#2\@@#3#4{\def#3{#1}\ifx #3\@nnil 
       \let\@nextwhile=\@psdonoop \else
      #4\relax\let\@nextwhile=\@tpsdoloop\fi\@nextwhile#2\@@#3{#4}}
\ifx\undefined\fbox
\newdimen\fboxrule
\newdimen\fboxsep
\newdimen\ps@tempdima
\newbox\ps@tempboxa
\long\def\fbox#1{\leavevmode\setbox\ps@tempboxa\hbox{#1}\ps@tempdima\fboxrule
    \advance\ps@tempdima \fboxsep \advance\ps@tempdima \dp\ps@tempboxa
   \hbox{\lower \ps@tempdima\hbox
  {\vbox{\hrule height \fboxrule
          \hbox{\vrule width \fboxrule \hskip\fboxsep
          \vbox{\vskip\fboxsep \box\ps@tempboxa\vskip\fboxsep}\hskip 
                 \fboxsep\vrule width \fboxrule}
                 \hrule height \fboxrule}}}}
\fi
\newread\ps@stream
\newif\ifnot@eof       % continue looking for the bounding box?
\newif\if@noisy        % report what you're making?
\newif\if@atend        % %%BoundingBox: has (at end) specification
\newif\if@psfile       % does this look like a PostScript file?
{\catcode`\%=12\global\gdef\epsf@start{%!}}
\def\epsf@PS{PS}
\def\epsf@getbb#1{%
\openin\ps@stream=#1
\ifeof\ps@stream\ps@typeout{Error, File #1 not found}\else
   {\not@eoftrue \chardef\other=12
    \def\do##1{\catcode`##1=\other}\dospecials \catcode`\ =10
    \loop
       \if@psfile
	  \read\ps@stream to \epsf@fileline
       \else{
	  \obeyspaces
          \read\ps@stream to \epsf@tmp\global\let\epsf@fileline\epsf@tmp}
       \fi
       \ifeof\ps@stream\not@eoffalse\else
       \if@psfile\else
       \expandafter\epsf@test\epsf@fileline:. \\%
       \fi
          \expandafter\epsf@aux\epsf@fileline:. \\%
       \fi
   \ifnot@eof\repeat
   }\closein\ps@stream\fi}%
\long\def\epsf@test#1#2#3:#4\\{\def\epsf@testit{#1#2}
			\ifx\epsf@testit\epsf@start\else
\ps@typeout{Warning! File does not start with `\epsf@start'.  It may not be a PostScript file.}
			\fi
			\@psfiletrue} % don't test after 1st line
{\catcode`\%=12\global\let\epsf@percent=%\global\def\epsf@bblit{%BoundingBox}}
\long\def\epsf@aux#1#2:#3\\{\ifx#1\epsf@percent
   \def\epsf@testit{#2}\ifx\epsf@testit\epsf@bblit
	\@atendfalse
        \epsf@atend #3 . \\%
	\if@atend	
	   \if@verbose{
		\ps@typeout{psfig: found `(atend)'; continuing search}
	   }\fi
        \else
        \epsf@grab #3 . . . \\%
        \not@eoffalse
        \global\no@bbfalse
        \fi
   \fi\fi}%
\def\epsf@grab #1 #2 #3 #4 #5\\{%
   \global\def\epsf@llx{#1}\ifx\epsf@llx\empty
      \epsf@grab #2 #3 #4 #5 .\\\else
   \global\def\epsf@lly{#2}%
   \global\def\epsf@urx{#3}\global\def\epsf@ury{#4}\fi}%
\def\epsf@atendlit{(atend)} 
\def\epsf@atend #1 #2 #3\\{%
   \def\epsf@tmp{#1}\ifx\epsf@tmp\empty
      \epsf@atend #2 #3 .\\\else
   \ifx\epsf@tmp\epsf@atendlit\@atendtrue\fi\fi}

\chardef\psletter = 11 % won't conflict with \begin{letter} now...
\chardef\other = 12

\newif \ifdebug %%% turn me on to see TeX hard at work ...
\newif\ifc@mpute %%% don't need to compute some values
\c@mputetrue % but assume that we do

\let\then = \relax
\def\r@dian{pt }
\let\r@dians = \r@dian
\let\dimensionless@nit = \r@dian
\let\dimensionless@nits = \dimensionless@nit
\def\internal@nit{sp }
\let\internal@nits = \internal@nit
\newif\ifstillc@nverging
\def \Mess@ge #1{\ifdebug \then \message {#1} \fi}

{ %%% Things that need abnormal catcodes %%%
	\catcode `\@ = \psletter
	\gdef \nodimen {\expandafter \n@dimen \the \dimen}
	\gdef \term #1 #2 #3%
	       {\edef \t@ {\the #1}%%% freeze parameter 1 (count, by value)
		\edef \t@@ {\expandafter \n@dimen \the #2\r@dian}%
		\t@rm {\t@} {\t@@} {#3}%
	       }
	\gdef \t@rm #1 #2 #3%
	       {{%
		\count 0 = 0
		\dimen 0 = 1 \dimensionless@nit
		\dimen 2 = #2\relax
		\Mess@ge {Calculating term #1 of \nodimen 2}%
		\loop
		\ifnum	\count 0 < #1
		\then	\advance \count 0 by 1
			\Mess@ge {Iteration \the \count 0 \space}%
			\Multiply \dimen 0 by {\dimen 2}%
			\Mess@ge {After multiplication, term = \nodimen 0}%
			\Divide \dimen 0 by {\count 0}%
			\Mess@ge {After division, term = \nodimen 0}%
		\repeat
		\Mess@ge {Final value for term #1 of 
				\nodimen 2 \space is \nodimen 0}%
		\xdef \Term {#3 = \nodimen 0 \r@dians}%
		\aftergroup \Term
	       }}
	\catcode `\p = \other
	\catcode `\t = \other
	\gdef \n@dimen #1pt{#1} %%% throw away the ``pt''
}

\def \Divide #1by #2{\divide #1 by #2} %%% just a synonym

\def \Multiply #1by #2%%% allows division of a dimen by a dimen
       {{%%% should really freeze parameter 2 (dimen, passed by value)
	\count 0 = #1\relax
	\count 2 = #2\relax
	\count 4 = 65536
	\Mess@ge {Before scaling, count 0 = \the \count 0 \space and
			count 2 = \the \count 2}%
	\ifnum	\count 0 > 32767 %%% do our best to avoid overflow
	\then	\divide \count 0 by 4
		\divide \count 4 by 4
	\else	\ifnum	\count 0 < -32767
		\then	\divide \count 0 by 4
			\divide \count 4 by 4
		\else
		\fi
	\fi
	\ifnum	\count 2 > 32767 %%% while retaining reasonable accuracy
	\then	\divide \count 2 by 4
		\divide \count 4 by 4
	\else	\ifnum	\count 2 < -32767
		\then	\divide \count 2 by 4
			\divide \count 4 by 4
		\else
		\fi
	\fi
	\multiply \count 0 by \count 2
	\divide \count 0 by \count 4
	\xdef \product {#1 = \the \count 0 \internal@nits}%
	\aftergroup \product
       }}

\def\r@duce{\ifdim\dimen0 > 90\r@dian \then   % sin(x+90) = sin(180-x)
		\multiply\dimen0 by -1
		\advance\dimen0 by 180\r@dian
		\r@duce
	    \else \ifdim\dimen0 < -90\r@dian \then  % sin(-x) = sin(360+x)
		\advance\dimen0 by 360\r@dian
		\r@duce
		\fi
	    \fi}

\def\Sine#1%
       {{%
	\dimen 0 = #1 \r@dian
	\r@duce
	\ifdim\dimen0 = -90\r@dian \then
	   \dimen4 = -1\r@dian
	   \c@mputefalse
	\fi
	\ifdim\dimen0 = 90\r@dian \then
	   \dimen4 = 1\r@dian
	   \c@mputefalse
	\fi
	\ifdim\dimen0 = 0\r@dian \then
	   \dimen4 = 0\r@dian
	   \c@mputefalse
	\fi
	\ifc@mpute \then
		\divide\dimen0 by 180
		\dimen0=3.141592654\dimen0
		\dimen 2 = 3.1415926535897963\r@dian %%% a well-known constant
		\divide\dimen 2 by 2 %%% we only deal with -pi/2 : pi/2
		\Mess@ge {Sin: calculating Sin of \nodimen 0}%
		\count 0 = 1 %%% see power-series expansion for sine
		\dimen 2 = 1 \r@dian %%% ditto
		\dimen 4 = 0 \r@dian %%% ditto
		\loop
			\ifnum	\dimen 2 = 0 %%% then we've done
			\then	\stillc@nvergingfalse 
			\else	\stillc@nvergingtrue
			\fi
			\ifstillc@nverging %%% then calculate next term
			\then	\term {\count 0} {\dimen 0} {\dimen 2}%
				\advance \count 0 by 2
				\count 2 = \count 0
				\divide \count 2 by 2
				\ifodd	\count 2 %%% signs alternate
				\then	\advance \dimen 4 by \dimen 2
				\else	\advance \dimen 4 by -\dimen 2
				\fi
		\repeat
	\fi		
			\xdef \sine {\nodimen 4}%
       }}

\def\Cosine#1{\ifx\sine\UnDefined\edef\Savesine{\relax}\else
		             \edef\Savesine{\sine}\fi
	{\dimen0=#1\r@dian\advance\dimen0 by 90\r@dian
	 \Sine{\nodimen 0}
	 \xdef\cosine{\sine}
	 \xdef\sine{\Savesine}}}	      
\def\psdraft{
	\def\@psdraft{0}
}
\def\psfull{
	\def\@psdraft{100}
}

\psfull

\newif\if@scalefirst
\def\psscalefirst{\@scalefirsttrue}
\def\psrotatefirst{\@scalefirstfalse}
\psrotatefirst

\newif\if@draftbox
\def\psnodraftbox{
	\@draftboxfalse
}
\def\psdraftbox{
	\@draftboxtrue
}
\@draftboxtrue

\newif\if@prologfile
\newif\if@postlogfile
\def\pssilent{
	\@noisyfalse
}
\def\psnoisy{
	\@noisytrue
}
\psnoisy
\newif\if@bbllx
\newif\if@bblly
\newif\if@bburx
\newif\if@bbury
\newif\if@height
\newif\if@width
\newif\if@rheight
\newif\if@rwidth
\newif\if@angle
\newif\if@clip
\newif\if@verbose
\newif\if@scale
\def\@p@@sclip#1{\@cliptrue}

\newif\if@decmpr

\def\@p@@sfigure#1{\def\@p@sfile{null}\def\@p@sbbfile{null}
	        \openin1=#1.bb
		\ifeof1\closein1
	        	\openin1=\figurepath#1.bb
			\ifeof1\closein1
			        \openin1=#1
				\ifeof1\closein1%
				       \openin1=\figurepath#1
					\ifeof1
					   \ps@typeout{Error, File #1 not found}
						\if@bbllx\if@bblly
				   		\if@bburx\if@bbury
			      				\def\@p@sfile{#1}%
			      				\def\@p@sbbfile{#1}%
							\@decmprfalse
				  	   	\fi\fi\fi\fi
					\else\closein1
				    		\def\@p@sfile{\figurepath#1}%
				    		\def\@p@sbbfile{\figurepath#1}%
						\@decmprfalse
	                       		\fi%
			 	\else\closein1%
					\def\@p@sfile{#1}
					\def\@p@sbbfile{#1}
					\@decmprfalse
			 	\fi
			\else
				\def\@p@sfile{\figurepath#1}
				\def\@p@sbbfile{\figurepath#1.bb}
				\@decmprtrue
			\fi
		\else
			\def\@p@sfile{#1}
			\def\@p@sbbfile{#1.bb}
			\@decmprtrue
		\fi}

\def\@p@@sfile#1{\@p@@sfigure{#1}}

\def\@p@@sbbllx#1{
		\@bbllxtrue
		\dimen100=#1
		\edef\@p@sbbllx{\number\dimen100}
}
\def\@p@@sbblly#1{
		\@bbllytrue
		\dimen100=#1
		\edef\@p@sbblly{\number\dimen100}
}
\def\@p@@sbburx#1{
		\@bburxtrue
		\dimen100=#1
		\edef\@p@sbburx{\number\dimen100}
}
\def\@p@@sbbury#1{
		\@bburytrue
		\dimen100=#1
		\edef\@p@sbbury{\number\dimen100}
}
\def\@p@@sheight#1{
		\@heighttrue
		\dimen100=#1
   		\edef\@p@sheight{\number\dimen100}
}
\def\@p@@swidth#1{
		\@widthtrue
		\dimen100=#1
		\edef\@p@swidth{\number\dimen100}
}
\def\@p@@srheight#1{
		\@rheighttrue
		\dimen100=#1
		\edef\@p@srheight{\number\dimen100}
}
\def\@p@@srwidth#1{
		\@rwidthtrue
		\dimen100=#1
		\edef\@p@srwidth{\number\dimen100}
}
\def\@p@@sangle#1{
		\@angletrue
		\edef\@p@sangle{#1} %\number\dimen100}
}
\def\@p@@srotate#1{\@p@@sangle{-#1}}
\def\@p@@sscale#1{
		\@scaletrue
		\edef\@p@sscale{#1}
}
\def\@p@@ssilent#1{ 
		\@verbosefalse
}
\def\@p@@sprolog#1{\@prologfiletrue\def\@prologfileval{#1}}
\def\@p@@spostlog#1{\@postlogfiletrue\def\@postlogfileval{#1}}
\def\@cs@name#1{\csname #1\endcsname}
\def\@setparms#1=#2,{\@cs@name{@p@@s#1}{#2}}
\def\ps@init@parms{
		\@bbllxfalse \@bbllyfalse
		\@bburxfalse \@bburyfalse
		\@heightfalse \@widthfalse
		\@rheightfalse \@rwidthfalse
		\@scalefalse
		\def\@p@sbbllx{}\def\@p@sbblly{}
		\def\@p@sbburx{}\def\@p@sbbury{}
		\def\@p@sheight{}\def\@p@swidth{}
		\def\@p@srheight{}\def\@p@srwidth{}
		\def\@p@sangle{0}
		\def\@p@sfile{} \def\@p@sbbfile{}
		\def\@p@scost{10}
		\def\@sc{}
		\@prologfilefalse
		\@postlogfilefalse
		\@clipfalse
		\if@noisy
			\@verbosetrue
		\else
			\@verbosefalse
		\fi
}
\def\parse@ps@parms#1{
	 	\@psdo\@psfiga:=#1\do
		   {\expandafter\@setparms\@psfiga,}}
\newif\ifno@bb
\def\bb@missing{
	\if@verbose{
		\ps@typeout{psfig: searching \@p@sbbfile \space  for bounding box}
	}\fi
	\no@bbtrue
	\epsf@getbb{\@p@sbbfile}
        \ifno@bb \else \bb@cull\epsf@llx\epsf@lly\epsf@urx\epsf@ury\fi
}	
\def\bb@cull#1#2#3#4{
	\dimen100=#1 bp\edef\@p@sbbllx{\number\dimen100}
	\dimen100=#2 bp\edef\@p@sbblly{\number\dimen100}
	\dimen100=#3 bp\edef\@p@sbburx{\number\dimen100}
	\dimen100=#4 bp\edef\@p@sbbury{\number\dimen100}
	\no@bbfalse
}
\newdimen\p@intvaluex
\newdimen\p@intvaluey
\def\rotate@#1#2{{\dimen0=#1 sp\dimen1=#2 sp
		  \global\p@intvaluex=\cosine\dimen0
		  \dimen3=\sine\dimen1
		  \global\advance\p@intvaluex by -\dimen3
		  \global\p@intvaluey=\sine\dimen0
		  \dimen3=\cosine\dimen1
		  \global\advance\p@intvaluey by \dimen3
		  }}
\def\compute@bb{
		\no@bbfalse
		\if@bbllx \else \no@bbtrue \fi
		\if@bblly \else \no@bbtrue \fi
		\if@bburx \else \no@bbtrue \fi
		\if@bbury \else \no@bbtrue \fi
		\ifno@bb \bb@missing \fi
		\ifno@bb \ps@typeout{FATAL ERROR: no bb supplied or found}
			\no-bb-error
		\fi
		\count203=\@p@sbburx
		\count204=\@p@sbbury
		\advance\count203 by -\@p@sbbllx
		\advance\count204 by -\@p@sbblly
		\edef\ps@bbw{\number\count203}
		\edef\ps@bbh{\number\count204}
		\if@angle 
			\Sine{\@p@sangle}\Cosine{\@p@sangle}
	        	{\dimen100=\maxdimen\xdef\r@p@sbbllx{\number\dimen100}
					    \xdef\r@p@sbblly{\number\dimen100}
			                    \xdef\r@p@sbburx{-\number\dimen100}
					    \xdef\r@p@sbbury{-\number\dimen100}}
                        \def\minmaxtest{
			   \ifnum\number\p@intvaluex<\r@p@sbbllx
			      \xdef\r@p@sbbllx{\number\p@intvaluex}\fi
			   \ifnum\number\p@intvaluex>\r@p@sbburx
			      \xdef\r@p@sbburx{\number\p@intvaluex}\fi
			   \ifnum\number\p@intvaluey<\r@p@sbblly
			      \xdef\r@p@sbblly{\number\p@intvaluey}\fi
			   \ifnum\number\p@intvaluey>\r@p@sbbury
			      \xdef\r@p@sbbury{\number\p@intvaluey}\fi
			   }
			\rotate@{\@p@sbbllx}{\@p@sbblly}
			\minmaxtest
			\rotate@{\@p@sbbllx}{\@p@sbbury}
			\minmaxtest
			\rotate@{\@p@sbburx}{\@p@sbblly}
			\minmaxtest
			\rotate@{\@p@sbburx}{\@p@sbbury}
			\minmaxtest
			\edef\@p@sbbllx{\r@p@sbbllx}\edef\@p@sbblly{\r@p@sbblly}
			\edef\@p@sbburx{\r@p@sbburx}\edef\@p@sbbury{\r@p@sbbury}
		\fi
		\count203=\@p@sbburx
		\count204=\@p@sbbury
		\advance\count203 by -\@p@sbbllx
		\advance\count204 by -\@p@sbblly
		\edef\@bbw{\number\count203}
		\edef\@bbh{\number\count204}
}
\def\in@hundreds#1#2#3{\count240=#2 \count241=#3
		     \count100=\count240	% 100 is first digit #2/#3
		     \divide\count100 by \count241
		     \count101=\count100
		     \multiply\count101 by \count241
		     \advance\count240 by -\count101
		     \multiply\count240 by 10
		     \count101=\count240	%101 is second digit of #2/#3
		     \divide\count101 by \count241
		     \count102=\count101
		     \multiply\count102 by \count241
		     \advance\count240 by -\count102
		     \multiply\count240 by 10
		     \count102=\count240	% 102 is the third digit
		     \divide\count102 by \count241
		     \count200=#1\count205=0
		     \count201=\count200
			\multiply\count201 by \count100
		 	\advance\count205 by \count201
		     \count201=\count200
			\divide\count201 by 10
			\multiply\count201 by \count101
			\advance\count205 by \count201
		     \count201=\count200
			\divide\count201 by 100
			\multiply\count201 by \count102
			\advance\count205 by \count201
		     \edef\@result{\number\count205}
}
\def\ps@scaleinhundreds#1{
		\in@hundreds{#1}{\@p@sscale}{100}
		\edef#1{\@result}
}
\def\compute@wfromh{
		\in@hundreds{\@p@sheight}{\@bbw}{\@bbh}
		\edef\@p@swidth{\@result}
}
\def\compute@hfromw{
	        \in@hundreds{\@p@swidth}{\@bbh}{\@bbw}
		\edef\@p@sheight{\@result}
}
\def\compute@handw{
		\if@height 
			\if@width
			\else
				\compute@wfromh
			\fi
		\else 
			\if@width
				\compute@hfromw
			\else
				\edef\@p@sheight{\@bbh}
				\edef\@p@swidth{\@bbw}
			\fi
		\fi
}
\def\compute@resv{
		\if@rheight \else \edef\@p@srheight{\@p@sheight} \fi
		\if@rwidth \else \edef\@p@srwidth{\@p@swidth} \fi
}
\def\compute@sizes{
	\compute@bb
	\if@scalefirst\if@angle
	\if@width
	   \in@hundreds{\@p@swidth}{\@bbw}{\ps@bbw}
	   \edef\@p@swidth{\@result}
	\fi
	\if@height
	   \in@hundreds{\@p@sheight}{\@bbh}{\ps@bbh}
	   \edef\@p@sheight{\@result}
	\fi
	\fi\fi
	\compute@handw
	\compute@resv
	\if@scale
	   \if@verbose
	      \ps@typeout{(scaling by \@p@sscale)}%
	   \fi
	   \ps@scaleinhundreds{\@p@swidth}%
	   \ps@scaleinhundreds{\@p@sheight}%
	   \ps@scaleinhundreds{\@p@srwidth}%
	   \ps@scaleinhundreds{\@p@srheight}%
	\fi
}

\def\psfig#1{\vbox {
	\ps@init@parms
	\parse@ps@parms{#1}
	\compute@sizes
	\ifnum\@p@scost<\@psdraft{
		\special{ps::[begin] 	\@p@swidth \space \@p@sheight \space
				\@p@sbbllx \space \@p@sbblly \space
				\@p@sbburx \space \@p@sbbury \space
				startTexFig \space }
		\if@angle
			\special {ps:: \@p@sangle \space rotate \space} 
		\fi
		\if@clip{
			\if@verbose{
				\ps@typeout{(clip)}
			}\fi
			\special{ps:: doclip \space }
		}\fi
		\if@prologfile
		    \special{ps: plotfile \@prologfileval \space } \fi
		\if@decmpr{
			\if@verbose{
				\ps@typeout{psfig: including \@p@sfile.Z \space }
			}\fi
			\special{ps: plotfile "`zcat \@p@sfile.Z" \space }
		}\else{
			\if@verbose{
				\ps@typeout{psfig: including \@p@sfile \space }
			}\fi
			\special{ps: plotfile \@p@sfile \space }
		}\fi
		\if@postlogfile
		    \special{ps: plotfile \@postlogfileval \space } \fi
		\special{ps::[end] endTexFig \space }
		\vbox to \@p@srheight true sp{
			\hbox to \@p@srwidth true sp{
				\hss
			}
		\vss
		}
	}\else{
		\if@draftbox{		
			\hbox{\frame{\vbox to \@p@srheight true sp{
			\vss
			\hbox to \@p@srwidth true sp{ \hss \@p@sfile \hss }
			\vss
			}}}
		}\else{
			\vbox to \@p@srheight true sp{
			\vss
			\hbox to \@p@srwidth true sp{\hss}
			\vss
			}
		}\fi

	}\fi
}}
\psfigRestoreAt
\let\@=\LaTeXAtSign